\documentclass{nature_fig}
    
    \usepackage{graphicx}
    \usepackage{ccaption}
    \usepackage{amsmath}
    \usepackage{amsfonts}
    \usepackage{amssymb}
    \usepackage{wasysym}
    \usepackage{xcolor}
    \usepackage{soul}
    \usepackage{lineno}
    \usepackage{gensymb}
    \usepackage{url}

    \def\be{\begin{equation}} 
    \def\ee{\end{equation}}

    \title{2025 Santorini-Amorgos crisis triggered by a transition from volcanic to regular tectonic activity}

    \author{Lippiello E.$^{1,*}$, Petrillo G.$^2$, Godano C.$^{1,3}$, Papadimitriou E. $^4$,  Karakostas V. $^4$ \&  Anagnostou V. $^4$}

\begin{document}
    
    \maketitle
    
    \begin{affiliations}
    \item [$^1$] Department of Mathematics and Physics, 
    University of Campania ``L. Vanvitelli'', 
    81100 Caserta, Italy.
    \item[$^2$] Earth Observatory of Singapore, Nanyang Technological University, Singapore
    \item [$^3$] Istituto Nazionale Geofisica e Vulcanologia, Napoli, Italy.
    \item[$^4$] Geophysics Department,
    Aristotle University of Thessaloniki,
    GR 541 24 Thessaloniki,
    Greece
    \end{affiliations}
    
\begin{abstract}
Fluid movement beneath volcanic regions can influence earthquake activity, but the processes linking seismic and volcanic systems are not fully understood. In early 2025, an unusual seismic sequence occurred close to Santorini, providing new insight into these interactions. Here we show that the sequence was likely initiated by the accumulation and migration of fluids beneath the volcanic complex. Seismic and ground deformation data reveal a progression from deep fluid buildup and microfracturing to the concentration of shallow earthquakes beneath Columbo volcano. This culminated in a four-day seismic episode that behaved like a single, slow-propagating rupture along a 16-kilometer fault, releasing energy equivalent to a magnitude 6.2 earthquake. The rupture was followed by a typical aftershock sequence. These observations suggest that fluid-driven processes can generate large earthquakes and redistribute stress in ways similar to tectonic mainshocks. This challenges conventional views on how seismic and volcanic hazards are connected and assessed.

\end{abstract}


\section{Introduction}
Fluid migration and stress redistribution play fundamental roles in driving both seismic and volcanic activity, particularly in tectonically active volcanic regions. Various mechanisms have been proposed to explain fluid-induced earthquakes, including pore-pressure diffusion and stress transfer across faults \cite{SKD13,dAGGL16,elsworth2016understanding,schultz2020hydraulic,shcherbakov2024stochastic}. These events are typically of small magnitude but may occur in swarms or clustered patterns. Similarly, changes in magma overpressure can promote the seismic triggering of volcanic eruptions through dike intrusion, chamber inflation, or hydrothermal activity \cite{MaBr06, SKWIJ21}. However, the interplay between these mechanisms and their broader implications for earthquake-volcano interactions remain open questions.

Santorini, located within the extensional Aegean back-arc region, provides an exceptional setting for investigating these coupled processes due to its complex tectonic environment and active volcanic system (see Supplementary Information). The region has experienced episodic seismic and volcanic activity over the past decades, with past sequences showing signs of both magmatic and tectonic influence.

In early 2025, a highly unusual seismic sequence occurred in the offshore area between Santorini and Amorgos Islands and the nearby Columbo volcano (see Fig. \ref{fig1}). The sequence did not begin with a single large rupture, but instead evolved gradually over several weeks. Here we analyze this sequence in detail, integrating ground deformation data, earthquake relocations, and refined statistical analysis of seismicity. We find evidence for a progression through five distinct phases, beginning with deep fluid accumulation and upward migration, followed by a slow rupture that released energy equivalent to a magnitude 6.2 earthquake, and culminating in a classical aftershock sequence. These findings reveal a scenario in which volcanic processes initiate tectonic-like seismicity, offering new insight into fluid-induced earthquake dynamics and challenging current models of seismic and volcanic hazard assessment.

\section{Results}

\subsection{\textbf{The past and present seismic activity.}}
\begin{figure}[h!]
        \centering
        \includegraphics[width=0.95\linewidth]{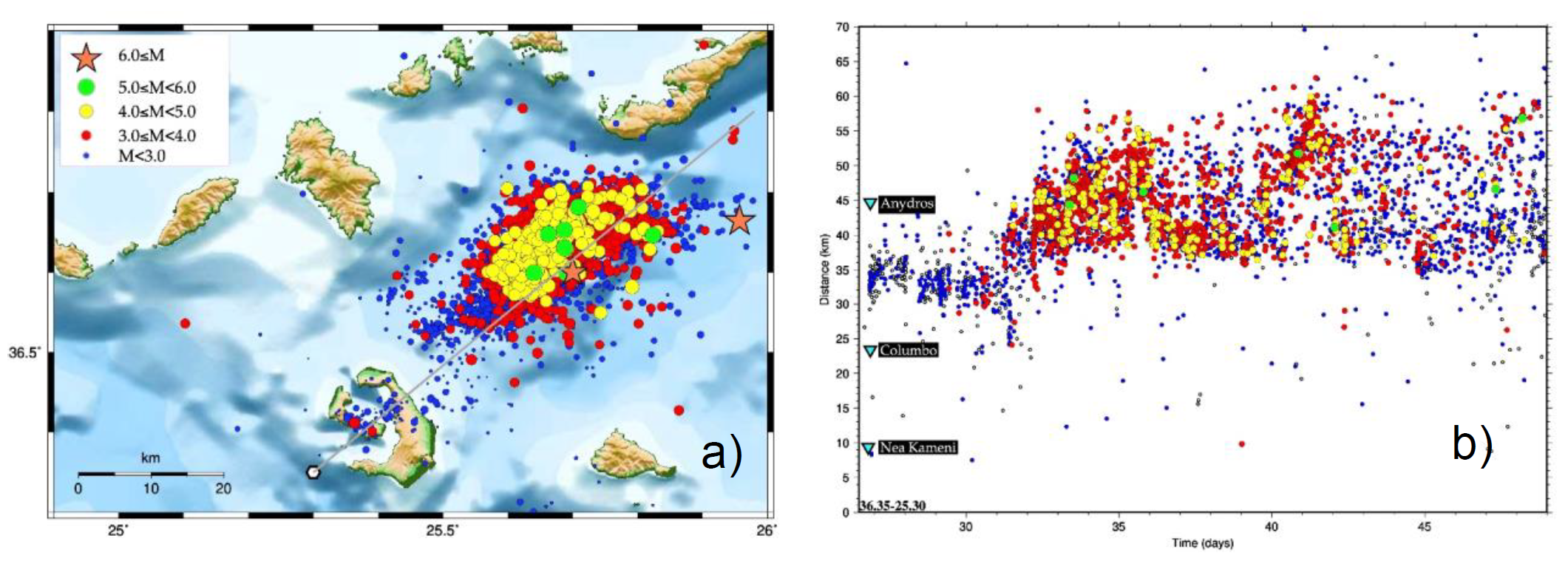}
        \caption{Seismicity evolution in the study area since 1 January 2025. (a) epicentral distribution of 6554 earthquakes with $M>1.0$ during January 1 – March 16, 2025. Symbol color and size are scaled according to earthquake magnitude, as shown in the legend. The two stars depict the 1956 main shocks, the bigger for the M7.7 Amorgos and the smaller for M6.9 Santorini earthquakes. The hexagon is the starting point since the space time plot is performed along the grey line. (b) Space time plot of the earthquakes shown on a map view in (a), starting on January 25 and lasting for 50 days. The locations of the Nea Kameni, Columbo and Anydros are shown by the inverse triangles with the corresponding location names to their next. Symbol color and size as in (a).}
        \label{fig1}
    \end{figure}

The earthquakes location in study area during the last 15 years, highlighting burst-like seismic activity near the Santorini Caldera (Suppl. Fig.2). A notable high number of earthquakes were recorded in the caldera during 2011 and 2012. This seismic sequence was followed by a quiescent period until 2024 when a clear reactivation was observed. This activity appears to be correlated with earthquakes close to the Columbo volcano. This area exhibited a significant increase of the seismic activity in 2011 and 2012. Here earthquakes occurrence persisted even during periods of low activity inside the caldera. Fig.\ref{fig1} shows the location map of the events recorded from January 2025.

In the following, we chronologically outline the five phases of tectonic activity that occurred between January 2024 and March 2025. For the analysis we used data from the relocated catalog\cite{KLAPAH25} that covers the period 2024/01/01 to 2025/03/16, comprising 6554 $M>1$ earthquakes.

\subsection{PHASE 1 -  Stationary regime}
\begin{figure}[h!]
        \centering
        \includegraphics[width=1.1\linewidth]{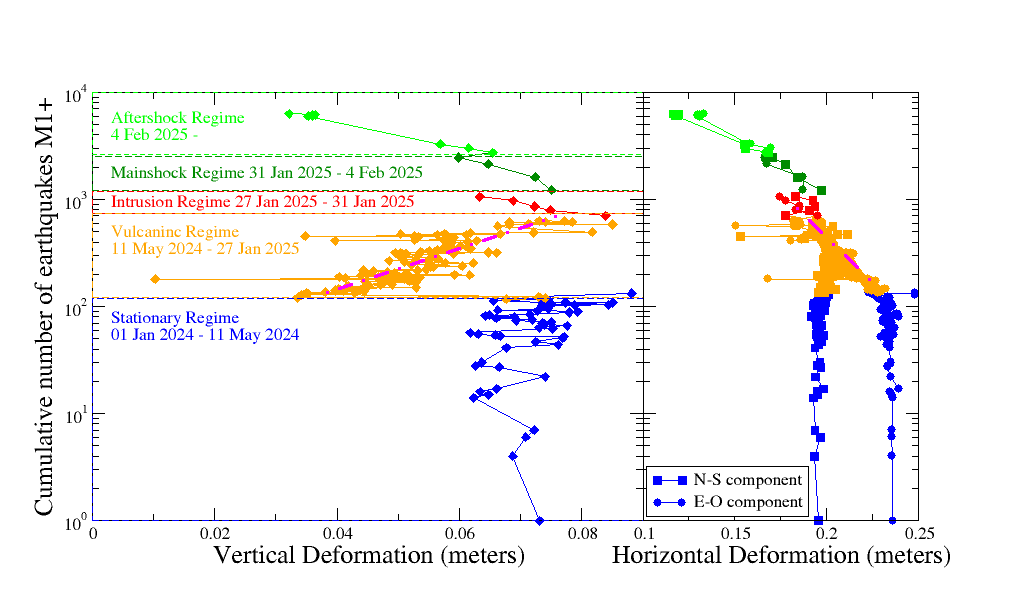}
\caption{The cumulative number $N(t)$ of recorded $M>1$ earthquakes, evaluated at time $t$ since January 1, 2024, is plotted against the deformation $\delta (t)$ measured at the continuous GPS station SANT at the same time $t$. (a) Vertical component. (b) Horizontal components, where the East-West component has been horizontally shifted to superimpose the North-South component. Different colors correspond to different phases. The magenta dashed line represents the exponential fit $N(t) \sim \exp{(c \delta(t))}$.}      
        \label{fig2}
    \end{figure}

Initially, from January 1 to May 11, 2024, seismicity in the Santorini region was comparative low and representative of typical stationary tectonic stress conditions. Seismicity spread out, predominantly reflecting normal faulting mechanisms consistent with the ongoing extension of the southern Aegean region.  
During this phase the ground deformation is roughly constant, besides some fluctuations, and this suggests that it is not the driving mechanism responsible for the observed seismic activity (Fig.\ref{fig2}). 

\subsection{PHASE 2 - Volcanic regime}
    The seismic regime exhibited a pronounced change on 11 May 2024, indicative of the onset of a distinct volcanic phase. The GPS SANT stations recorded significant vertical deformation, signaling substantial changes in subsurface stress conditions. In fact, plotting the cumulative number of earthquakes as a function of the ground deformation (see methods), we observe a clear exponential trend (Fig. \ref{fig2}). This is indicative of a mechanism of earthquake occurrence driven by the ground deformation caused by an enhanced pressurization from depth \cite{voi88,VC91,kle00,Kil12,KDC17,KCDP23,BNDGMR24,GCTP25}. An exponential trend is also observed in the East-West GPS component, consistent with deformation influencing seismicity, while the North-South component still appears constant showing no correlation with earthquake occurrence. Concurrently, seismicity developed into distinct swarms concentrated initially beneath Santorini caldera, subsequently migrating progressively toward Columbo volcano (see Supp. Fig.3). In total, we recorded 552 earthquakes with $M > 1$, with a maximum magnitude of $M = 3.8$ \cite{karakostas2025santorini}. The overall picture in this phase     
    are consistent with an increase of magmatic overpressure, commonly preceding volcanic eruptions or significant magmatic intrusions. \\

\begin{figure}
    \centering
    \includegraphics[width=0.9\linewidth]{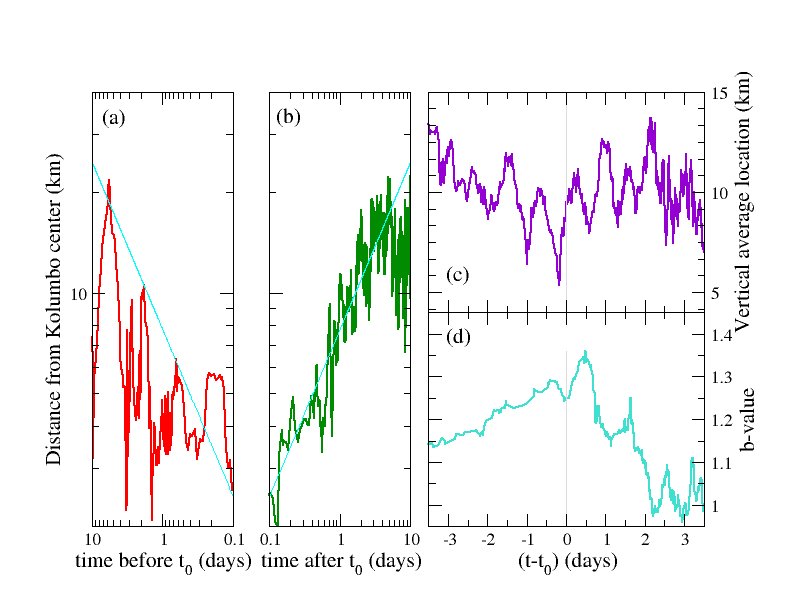}
    \caption{The evolution of seismicity during the Intrusion and Mainshock phases. (a) The average epicentral distance $\delta r(t)$ from the Columbo volcano center $\mathbf{x_K}$ as a function of the time difference from $t_0$ (12:00 on January 31, 2025) for times before $t_0$. (b) Same as panel (a), but for times after $t_0$. The turquoise line represents the fit $\delta r(t)=\sqrt{4 \pi D |t-t_0|}$ \cite{shapiro1997estimating}, expected for a diffusion process with a diffusion coefficient $D=4.8 \, \text{km}^2/\text{day}$. (c) The average focal depth of $M>1$ earthquakes as a function of $t-t_0$. (d) The b-value, estimated using the b-more-positive method, is plotted as a function of $t-t_0$.}
    \label{fig3}
\end{figure}

\subsection{PHASE 3 -  Intrusion regime}
Between January 27 and January 31, 2025, seismicity underwent a distinct transition phase. We define the reference time $t_0$ as 12:00 on January 31, 2025. The four days leading up to this point exhibit clear indicators of a transition phase, characterized by several marked observations:
\begin{itemize}
\item A change in the deformation directionality of the vertical GPS component, signaling the onset of downward deformation at the SANT station location (Fig.\ref{fig2}).
\item Convergence of seismic activity towards the center of the Columbo volcano, $\mathbf{x_k}=(36.537,25.547)$, as evident from the average epicentral distance from $\mathbf{x_K}$ (Fig.\ref{fig3}a).
\item A decrease in the average focal depth of recorded seismicity, from approximately 15 km to about 5 km as $t_0$ is approached (Fig.\ref{fig3}c).
\item An increase in the Gutenberg-Richter $b$-value, calculated using the $b$-more-positive method \cite{VdE21,LP24} (see methods), peaking at $b \simeq 1.4$ shortly after $t_0$ (Fig.\ref{fig3}d).
\end{itemize}
These combined results suggest the ascent of magmatic fluids, pressurizing shallower reservoirs located beneath the Columbo volcano. During this transition phase, 378 earthquakes with $M > 1$ were located, reaching a maximum magnitude of $M = 3.4$.

\subsection{PHASE 4 - Mainshock regime} Remarkably, starting January 31, seismicity transitioned into an unusual phase characterized by a sharp increase of the seismic activity, with all the patterns identified in the previous phase — such as convergence towards $\mathbf{x_k}$, a decrease in focal depth, and an increase in the $b$-value — exhibiting a sudden reversal (see Fig.~\ref{fig3}).
 During the subsequent 4 days after $t_0$, we located 1401 earthquakes with $M > 1$, including approximately 20 events with magnitudes exceeding $M = 4.5$. Due to short-term aftershock incompleteness, the actual number of $M > 1$ earthquakes, in this period is likely significantly higher than the recorded count. An estimate using the $a$-positive method \cite{VdEP23}(see Methods) suggests about $3 \times 10^6$ $M > 1$ earthquakes, with a maximum magnitude $M_{sup}=5.1$ and a cumulative seismic moment released comparable to a single magnitude $M \simeq 6.2$ earthquake (see Methods).

This phase (see Fig.~\ref{fig3}b) exhibits sustained seismic activity that diffusively spread along fault structures, over the four days. The estimated seismic diffusivity is $D \approx 4.8 \, \text{km}^2/\text{day}$ (Fig.\ref{fig3}b), which strongly supports a fluid-driven mechanism \cite{petrillo2024fluids, gentili2024seismic}. In this scenario, magmatic fluids likely exploited the pre-existing fracture networks formed during the preceding microfracturing phase, permeating adjacent tectonic faults.

In our framework, the seismic activity can be interpreted as a single large mainshock with a propagating front-wave that activated a fault segment approximately $16$ km in length, which is the typical length \cite{WC94} of an approximately $M \simeq 6.3$ earthquake. Unlike typical tectonic earthquakes, which are characterized by brief and intense rupture propagation, lasting some seconds, this fluid-driven mainshock lasted a few days.

The front-wave propagation was accompanied by deformation that aligned along the fault direction (approximately 45 degrees from the SANT station), as evidenced by the nearly parallel evolution of the horizontal deformation components (Fig.\ref{fig2}b). Simultaneously, the vertical deformation showed a continuous decrease.

\begin{figure}
    \centering
    \includegraphics[width=0.9\linewidth]{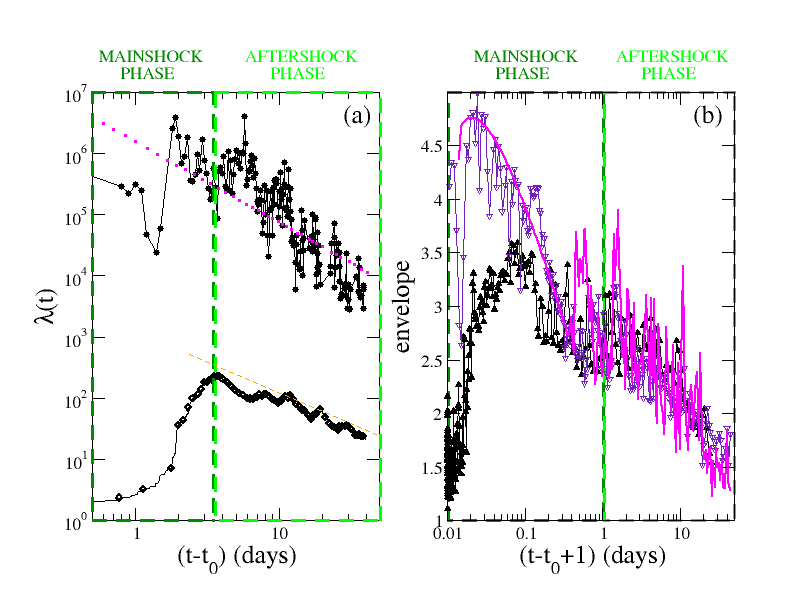}
\caption{Seismic activity in the aftershock regime. (a) The daily occurrence rate of located earthquakes with $M>1$ is plotted as a function of time, with $\lambda(t)$ estimated using traditional methods (open diamonds) and the a-positive method (filled circles). (b) The envelope function (see Methods) of the signal recorded at the Thera stations is plotted as a function of time since February 2, 2025 (filled upward triangles). Empty violet triangles represent the envelope of the signal recorded at the DAT station for the $M=6.1$ Kos earthquake, vertically shifted by subtracting $1.8$ and with the time axis scaled by a factor of $300$. The magenta dashed line represents the synthetic envelope expected for a regular aftershock sequence triggered by an $M>6$ mainshock, with the time scale similarly scaled by a factor of $300$ (see Methods).}  
    \label{fig4}
\end{figure}

\subsection{PHASE 5 - Aftershock regime}
Following February 4, seismic activity transitioned into a well-defined aftershock regime, characterized by a systematic temporal decay in seismicity rates accurately described by the Omori-Utsu empirical law \cite{utsu1995centenary} (Fig.\ref{fig4}). Quantitative analysis of this decay revealed an Omori exponent $p \approx 1.5$, significantly exceeding typical tectonic aftershock sequences, which generally exhibit exponents in the range $p = 0.8 - 1.2$. Elevated $p$-values are consistent with previous observations of fluid-induced seismicity, reflecting accelerated stress relaxation facilitated by rapid pore-pressure diffusion \cite{langenbruch2010decay,farahbod2019aftershock}. \\
The physical mechanism underlying this accelerated decay is related to fluid-pressure dynamics. During the preceding seismic phase, magmatic fluid injection substantially elevated pore-fluid pressures, reducing effective normal stresses and enabling widespread fracture reactivation. After the cessation of fluid injection, rapid diffusion of pore pressures occurred, swiftly re-establishing higher effective stresses and causing a rapid decrease of seismicity. This fluid-driven pore-pressure diffusion mechanism inherently promotes faster stress equilibration, yielding elevated Omori decay rates $(p > 1.2)$, compared to purely seismic sequences \cite{langenbruch2010decay}.

The idea of a behavior consistent with a ``classical'' aftershock regime is confirmed by the behavior of the envelope function $\mu(t)$ (Fig.4b). The temporal evolution of $\mu(t)$ is in fact closely related to the temporal organization of the aftershock in seismic sequences \cite{LCGPK16,LCGPK19,LPGTPK19} (Methods). To support our conclusion we compare $\mu(t)$ obtained from the signal recorded at the THERA station with the one obtained for the KOS $M=6.1$ earthquake in 2017 which is the closest in space $M>6$ mainshock in the Aegean, involving a fault segment similar to the one involved by the Santorini-Amorgos sequence. More precisely, we find that by multiplying the time scale by a factor $300$, the decay of $\mu(t)$ for Santorini-Amorgos and Kos sequences is similar during the aftershock phase ($t>t_0+4$). This rescaling of the time axis reflects the difference in the time scale between the two sequences with processes during the Santorini-Amorgos sequence abnormally slower than classical main-aftershock activity. A key difference remains in the envelope during the mainshock phase ($ t < t_0 + 4 $): the peak value of $\mu(t)$ is significantly lower for the Santorini-Amorgos sequence compared to the Kos sequence. This can be readily explained by the anomalous nature of the Santorini-Amorgos mainshock, which unfolds over a much longer time window than the Kos event. However, as previously mentioned, the total seismic energy released during the mainshock phase is comparable in both cases.

\section{Discussion and Conclusion}
    This unusual seismic sequence underscores several critical scientific implications. Firstly, it demonstrates how seismic events may effectively mitigate volcanic eruption hazards by relieving magmatic overpressure through tectonically facilitated fluid migration. The extensive and rapid opening of fracture networks, as evidenced by high $b-$values and the vertical deformation reversal, likely prevented what could have evolved into an eruptive scenario. Secondly, the prolonged and fluid-driven mainshock-like seismicity challenges traditional seismic hazard assessments, calling for revised models that incorporate fluid migration and stress redistribution over extended timescales. Lastly, this sequence reveals the necessity of integrating both volcanic and tectonic components into hazard assessments, especially in volcanic arcs with active tectonic fault systems. \\
    Ultimately, the Santorini-Amorgos seismic sequence of February 2025 provides a remarkable natural example of how fluid-driven seismicity can modulate volcanic hazards, disrupt traditional seismic paradigms, and reshape our understanding of earthquake sequences. The comprehensive analysis of deformation data, statistical parameters (high $b-$values and elevated $p-$values), depth migration, and fluid diffusion processes collectively supports a unified interpretation that significantly advances the scientific understanding of volcano-tectonic interactions. This case study provides essential insights for developing future hazard mitigation strategies globally, especially in similarly complex volcanic-tectonic settings.


\section{Methods}
\subsection{Relocation Procedure}
The relocated earthquake catalog of NLL-SSST-coherence was based on phase arrival data taken from the bulletins of the Aristotle University of Thessaloniki, Geophysics Department \cite{AUT81} and the National Observatory of Athens, Institute of Geodynamics \cite{NOA75}. The NLL-SSST-coherence procedure \cite{LoSa22,LoHe23} was applied for the relocation and the seismic velocity model was from \cite{PKKKVM15}.

\subsection{The plot of cumulative number of events versus ground deformation.}
We analyze the daily ground deformation $\delta(t)$ recorded at station SANT, located on Santorini Island (coordinates 36.4336N 25.4226E), starting from January 1, 2024. Over the same period, we measure the cumulative number of relocated earthquakes with magnitude $M \geq 1$, denoted as $N(t)$. The parametric plots in Fig. \ref{fig2} are obtained by plotting $N(t)$ against $\delta(t)$ at the same time $t$, considering the three deformation components. Similar results are obtained by considering for $N(t)$ the cumulative number of $M > 2.5$ earthquakes.

\subsection{Estimation of the $b$ value.}

The $b$-value, which determines the slope of the earthquake frequency–magnitude distribution, according to the Gutenberg-Richter (GR) law $P(m) = b \log(10) 10^{-bm}$,  is a crucial parameter in seismic forecasting \cite{GR44}. To overcome issues related to temporal and spatial variations in the completeness of seismic catalogs \cite{Kag04,HKJ06,PVIH07,Hai16,Hai16a,dAGL18,Hai21}, we estimate the $b$-value using the b-more-positive estimator \cite{LP24}, which relies on positive magnitude difference statistics \cite{VdE21}.
Specifically, for each earthquake $i$ with magnitude $m_i$, we identify the first subsequent earthquake $j$ with magnitude $m_j > m_i + \delta M_{th}$, occurring within one day and within $25$ km of the epicenter of event $i$. We then compute the magnitude difference as $\delta m_i = m_j - m_i$. Focusing only on earthquakes followed by such an event, we calculate the average value of $\delta m_i$ for this subset and estimate the $b$-value as $b=\frac{1}{\log(10) \langle \delta m \rangle}$. Results are obtained by setting $\delta M_{th}=0.5$ and applying a smoothing procedure on overlapping windows including $10$ events. Similar results are found for $\delta M_{th} \in [0.4,0.7]$.

\subsection{The diffusion from the Columbo vulcano center}
We compute the epicentral distance of all earthquakes from the coordinates $\mathbf{x_k}=(36.537,25.547)$ identified as the center of the Columbo volcano. The average epicentral distance $\langle \Delta r \rangle$ is obtained by averaging this distance over overlapping windows of $10$ events. The same sliding average procedure is applied to measure the average focal depth as a function of time (Fig. \ref{fig3}c). The diffusion coefficient $D$ is estimated by fitting the relation $\langle \Delta r(t) \rangle=\sqrt{4\pi (t-t_0)}$.

\subsection{Estimate of the equivalent moment magnitude during the mainshock phase}
The total seismic moment release, $M_{tot} = \sum_{i=1}^N 10^{\frac{3}{2}m_i} = N b \log(10) \int_1^{M_{sup}}  dm  \; 10^{-bm}10^{\frac{3}{2}m} $, where $N$ is the total number of $M>1$ earthquakes and we assume that they follow the GR law up to an upper magnitude $M_{sup}$. By setting $N=3 \times 10^6$ and $M_{sup}=5.1$, and an average value $b=1$, obtained from the experimental observations, we obtain $M_{tot}\simeq2\times10^9$ corresponding to a single earthquake of magnitude $6.2$.

\subsection{Evaluation of the seismic occurrence rate $\Lambda(t)$.}

The traditional estimate of $\Lambda(t)$ is obtained by measuring the time $\tau_k$ required to observe $k=200$ events, yielding $\Lambda(t)=k/\tau_k$.
The a+ approach \cite{VdEP23} estimates the occurrence rate $\Lambda_+(t)$ using the time interval $\delta t_{ij}$ between the same earthquake pairs considered in the evaluation of the $b$-value, according to the b-more-positive estimator (see above). Specifically, for each earthquake $i$ included in the $b$-value calculation, we also evaluate $\lambda_i=10^{b (m_i+\delta M_{th})}/\delta t_{ij}$, which is the expected occurrence rate in the considered time window \cite{}. The quantity $\Lambda_+(t)$ is then given by $\Lambda_+(t)=\frac{1}{k} \sum_{i=1}^k(1/\lambda_i)$,
where the sum runs over $k=200$ events in the interval starting at time $t$. We average $1/\lambda_i$ instead of $\lambda_i$ to reduce fluctuations \cite{VdEP23}.

\subsection{The envelope of the seismic signal.}
The vertical component of the ground velocity is processed by applying a band-pass filter in the frequency range $[2,10]$ Hz, followed by a Hilbert transform. We then compute the logarithm of the resulting signal.
The envelope function $\overline{\mu}(t)$ is obtained through a logarithmic smoothing procedure, averaging over windows of increasing duration
$\Delta t_k=\Delta t_0 (1+\zeta)^k$ with $\Delta t_0=0.1$ sec and $\zeta=0.005$.

The generation of the simulated signal follows the routine presented in  \cite{LPGTPK19}. Specifically, we assign a mainshock of magnitude $M=6$ at time $t=0$ and generate $N_{\text{aft}}$ aftershocks distributed in time according to the Omori-Utsu law. We set $N_{\text{aft}} = 375000$ to reflect the typical number of aftershocks with $M \geq 1$ triggered by an $M>6$ mainshock.

We assume that aftershocks follow the GR law with a $b$-value of $b=0.95$ and associate each aftershock with an envelope based on the power-law functional form proposed in  \cite{LCGPK16,LCGPK19,LPGTPK19}. The synthetic envelope is then obtained as the superposition of the individual aftershock envelopes.
In the synthetic envelope shown in Fig. \ref{fig4}b, time scales are multiplied by a factor of $300$ compared to typical values used to simulate normal aftershock sequences.

\section*{Data Availability}
The datasets generated and analyzed during this study is available at https://doi.org/10.5281/zenodo.15111649.

\section*{Author Contributions}
All authors have contributed equally to conceptual development, drafting, and revision of the manuscript. 

\section*{Competing Interests}
The authors declare no competing interests.

\section*{Acknowledgments}
 E.L. acknowledges support from the MIUR PRIN 2022 PNRR  P202247YKL,
  C.G. acknowledges support from the MIUR PRIN 2022 PNRR P20222B5P9.

\section*{References}

\bibliography{biblio}
    
\end{document}